\documentclass[letter]{aa} % for the letters 
\usepackage{graphicx}
\usepackage{txfonts}
\usepackage{natbib} 
\begin{document}
%
   %\title{Explaining the peculiar X-ray emission of CTTS}
   \title{Correlated optical and X-ray variability in CTTS}

   %\subtitle{Simultaneous CoRoT and {\em Chandra} observations of NGC~2264}
   \subtitle{Indications of absorption-modulated emission}

\author{E. Flaccomio\inst{1} \and G. Micela\inst{1} \and F. Favata\inst{2} \and S.P.H. Alencar\inst{3}}
\institute{INAF - Osservatorio Astronomico di Palermo, 
  Piazza del Parlamento 1, I-90134 Palermo, Italy \\ \email{E. Flaccomio, ettoref@astropa.unipa.it}
 \and
 European Space Agency, 8-10 rue Mario Nikis, 75015 Paris, France
 \and
 Departamento de Fisica - ICEx - UFMG, Av. Ant\^onio Carlos, 6627, 30270-901, Belo Horizonte, MG, Brazil
}

\date{Received May 3, 2010; accepted May 25, 1010}

% \abstract{}{}{}{}{} 
% 5 {} token are mandatory
 
  \abstract
   {}
   {Optical and X-ray emission from classical T Tauri stars (CTTSs) has long been known to be highly variable. Our long, uninterrupted optical observation of the NGC\,2264 region with CoRoT\thanks{ The CoRoT  space mission was developed and is operated by the French space agency CNES, with participation of ESA's RSSD and Science Programs, Austria, Belgium, Brazil, Germany, and Spain} allows the optical variability in CTTS to be studied with unprecedented accuracy and time coverage. Two short Chandra observations obtained during the CoRoT pointing with a separation of 16 days allow us to study whether there is a correlation between optical and X-ray variability on this timescale, thus probing the physical mechanisms driving the variability in both bands.}
   {We have computed the optical and X-ray fractional variability between the two 30 ks duration windows covered by both the Chandra and CoRoT observations, for a sample of classical and weak line T Tauri stars (WTTSs) in NGC\,2264. A scatter plot clearly shows that the variability of CTTSs in the optical and soft X-ray (0.5-1.5\,keV) bands is correlated, while no correlation is apparent in the hard (1.5-8.0\,keV) band. Also, no correlation in either band is present for WTTSs.}
   {We show that the correlation between soft X-ray and optical variability of CTTSs can be naturally explained in terms of time-variable shading (absorption) from circumstellar material orbiting the star, in a scenario rather similar to the one invoked to explain the observed phenomenology in the CTTS AA\,Tau. The slope of the observed correlation implies (in the hypothesis of homogeneous shading) a significant dust depletion in the circumstellar material (with a gas-to-dust ratio approximately 5 times lower than the standard value for interstellar material).}
   {}

   \keywords{Stars: variable: T Tauri, Herbig Ae/Be -- Stars: activity -- Stars: coronae -- Stars: formation -- Accretion, accretion disks -- X-rays: stars}

   \maketitle
%
%________________________________________________________________

\section{Introduction}

Strong irregular optical variability is one of the defining characteristics of classical T-Tauri stars \citep[CTTS,][]{joy45}, together with their association with dark or bright nebulae and their strong optical emission lines. Following their discovery by \citet{joy45}, CTTS were then recognized as newly formed stars that have just completed their main accretion phase and are contracting toward the main sequence \citep[e.g.][]{wal56}.
Unlike weak line T-Tauri stars (WTTS), CTTS are still undergoing mass-accretion: material from the inner edge of their truncated circumstellar disk is thought to be channeled along magnetic field lines toward an impact region on the stellar surface where it is shocked, producing an excess of emission with respect to the stellar photosphere, ranging from the X-ray band to the optical. The large and irregular optical variability of CTTSs has generally been linked with the accretion process, but the actual mechanism involved has remained elusive, with two classes of mechanisms being generally considered, i.e., $i$) variability of the emission from the accretion shock(s) due to variation in mass accretion rate and/or to their rotational modulation, or $ii$) variable absorption due to unstable and optically thick accretion streams and/or warps in the circumstellar disk that occult part of the photosphere. Recent statistical studies of CTTS observed over many years \citep{gra07,gra08} suggest that in about 25\% of the cases the optical variability may be attributed to absorption, while in the remaining cases time-variable accretion is favored. However, due to the similar effects of spots and absorption on the broad-band lightcurves, the absorption scenario cannot be excluded in most cases. A detailed study of the CTTS AA\,Tau by Bouvier and collaborators \citep{bou99,bou03,men03,bou07,gro07} has shown that, for this near edge-on star-disk system, the large, irregular optical variability is explained well by occultation of a significant fraction of the stellar surface by a warp in the inner disk, located at the corotation radius and thus rotating in and out of view with the same period as the stellar photosphere and evolving on similar timescales. The warp could be due to the misalignment of the rotation and magnetic axes and could correspond to the foot of the accretion stream. Until very recently the case of AA Tau could be considered peculiar, as no other similar systems had been reported. However, recent high-quality optical lightcurves of a large sample of young stars, CTTS and WTTS, in the NGC\,2264 star-forming region, obtained with the CoRoT satellite, have shown that AA\,Tau-like variability is rather common \citep{ale10}. This leads to the suggestion that time-dependent obscuration of part of the photosphere by disk warps, or else by the related accretion streams, might be an important mechanism to explain the optical variability of CTTS.

CTTS are also peculiar in the X-ray band; like WTTS, they show significant coronal emission from plasma at 10-30\,MK, but with average luminosities lower than those of WTTSs, at any given stellar mass or bolometric luminosity, by a factor of 3-5, and with a significantly larger scatter \citep{fla03b,pre05a}. Moreover, the X-ray emission of CTTSs may be more time variable and have a harder spectrum than the one of WTTSs \citep[e.g.][]{ima01,fla06}. These facts have so far remained unexplained. A variety of physical mechanisms have been suggested to explain the lower observed X-ray luminosity, such as mass-loading of coronal magnetic loops due to accretion material (leading to cooler plasma, not visible in X-rays) and the shielding of significant fractions of the coronal plasma by dense accretion streams \citep{gre07}.

In addition to the {\em hard} coronal emission, a separate 2-3 MK X-ray spectral component related to accretion and possibly originating in the accretion shock is now believed to be common in CTTSs. This soft component has only been observed to date in a dozen such stars observed at high spectral resolution with either {\em Chandra} or {\em XMM-Newton} \citep[e.g.][]{gue07b}. With the notable exception of TW\,Hya, the coronal component seems to dominate the emission for $E>500$\,eV. 

Simultaneous optical and X-ray observations can constrain the physical mechanisms responsible for the optical and X-ray variability and the location of the X-ray emitting material relative to the photosphere. For the case of AA\,Tau, for example, \citet{gro07} searched for X-ray eclipses corresponding to two of the optical eclipses caused by the disk warp. Their failure to detect X-ray eclipses was interpreted as evidence that the X-ray emitting plasma is at high latitudes. \citet{sta06} and \citet{sta07} examined ground-based optical photometry of $\sim 800$ young stars in the Orion Nebula Cluster, overlapping, for $\sim 1$ week, with the {\em Chandra} Orion Ultradeep Project (COUP) observation of the same region. They found ``very little evidence to suggest a direct causal link between the sources of optical and X-ray variability in PMS stars''.

We have obtained two $\sim$30\,ksec {\em Chandra} observations of the star-forming region NGC\,2264 overlapping with a dedicated CoRoT ``short run''. The two {\em Chandra} observations, separated by approximately two weeks, were allocated from the Director's Discretionary Time. We present evidence that the soft X-ray and optical emissions are correlated for CTTSs (but not for WTTSs) and discuss the implications in terms of location of the X-ray emitting material and origins of the variability.

\section{Data and sample selection}
\label{sect:data}
 
We observed the star-forming region NGC~2264 for 23.5 days with the CoRoT satellite in March 2008. The two CCDs normally used for exoplanet observations cover a $\sim$2 sq.degree field with the cluster fitting in a single CCD. High-quality, broadband (370-950\,$\mu$m),  optical lightcurves were obtained with a cadence of 512 or 32 seconds for 8150 pre-selected targets in the field, with magnitudes down to $I\sim16$. First results for NGC~2264 members were published by \citet{ale10}, while a complete description of the observation is in preparation (Favata et al.). We here use CoRoT ``white light'' lightcurves, as produced by the standard pipeline, cleaned of datapoints of dubious quality (status flag $\ne$ 0) and all rebinned to 512 seconds. The time series for $\sim1/3$ of the stars in our final sample actually contain color information. We, however, decided to use only the sum of the fluxes in the three available bands, given the poor definition of the CoRoT photometric system.

During the CoRoT pointing, we obtained two {\em Chandra} ACIS-I observations of a $\sim$$17\arcmin\times17\arcmin$ field in NGC~2264, the first on 12 March, lasting 28\,ksec (ObsId: 9768), and the second on 28 March, lasting 30\,ksec (ObsId: 9769). The aim points were the same within 2\arcsec  (R.A. 6:41:12.5, Dec. +9:29:32) and the roll angles differed by only $\sim4$ degrees, so that the two fields overlap almost completely. 
A full account of the analysis of the {\em Chandra} observations will be provided by Flaccomio et al. (in preparation). In brief, we performed source detection on each field with PWdetect \citep{dam97} and then used {\em ACIS Extract} \citep{bro10} to extract individual source spectra and lightcurves. We here make use mainly of the mean observed fluxes during the two observations, in units of counts~s$^{-1}$cm$^{-2}$, i.e., countrates divided by the effective area of the detector at each source position.

In the following we study the optical and X-ray flux variations between the times of the two {\em Chandra} pointings. The reference source sample was defined starting with the 81 sources in common between the CoRoT and the {\em Chandra} datasets and with unambiguous cross-identifications with optical and NIR catalogs \citep[][2MASS]{sun08,sun09}. We then excluded five stars whose CoRoT lighturves are affected by sudden flux variations, most likely spurious and due to cosmic rays. In the X-ray band we consider two different energy ranges: 0.5-1.5\,keV (soft) and 1.5-8.0\,keV (hard). To limit the uncertainties on the flux differences between the two observations we restricted our sample to stars with $\geq$5 X-ray photons detected in the band of interest in {\em at least one} of the two observations. This condition leads to samples of 69 and  62 objects for the soft and hard X-ray bands, respectively. Finally, we excluded  14 objects from both samples: three are not low-mass stars, but Herbig Ae/Be stars, based on their spectral types and/or V-I colors; the other 11 showed evidence of strong X-ray flares during the {\em Chandra} observations. Although of interest, flares were excluded for the present investigation since our aim here is to investigate flux variability due either to accretion or to variable absorption, i.e. mechanisms unrelated to flaring activity. Our main conclusions, however, are hardly affected by the exclusion of these 11 stars. The final samples of stars with ``good'' simultaneous optical and X-ray fluxes count 55 and 48 objects considering the soft and hard X-ray bands, respectively. The smaller sample is, with the exception of one star, a subset of the larger one.

We finally classify the stars in our sample as CTTSs or WTTSs following
\citet{ale10}: CTTSs are defined as stars with an H$_\alpha$ equivalent
width $>$$10$\,$\AA$, or H$_\alpha$ width at 10\% of the peak intensity
greater than 270\,km\,s$^{-1}$ \citep[data
from][]{dah05,fur06}\footnote{A third criterium used by \citet{ale10},
based on the $U-V$ excess, was redundant for our smaller sample.}. The
other stars were classified as WTTSs. Like CTTS, all WTTS in our sample
are almost certainly members of NGC~2264, being detected in X-rays and
satisfying, in the vast majority of cases, other membership criteria
based on radial velocity, optical variability, and position in optical
color-magnitude diagrams \citep{fur06,lam04}.

\section{Results}

\begin{figure*}[!th!]
\centering
\includegraphics[width=9.15cm]{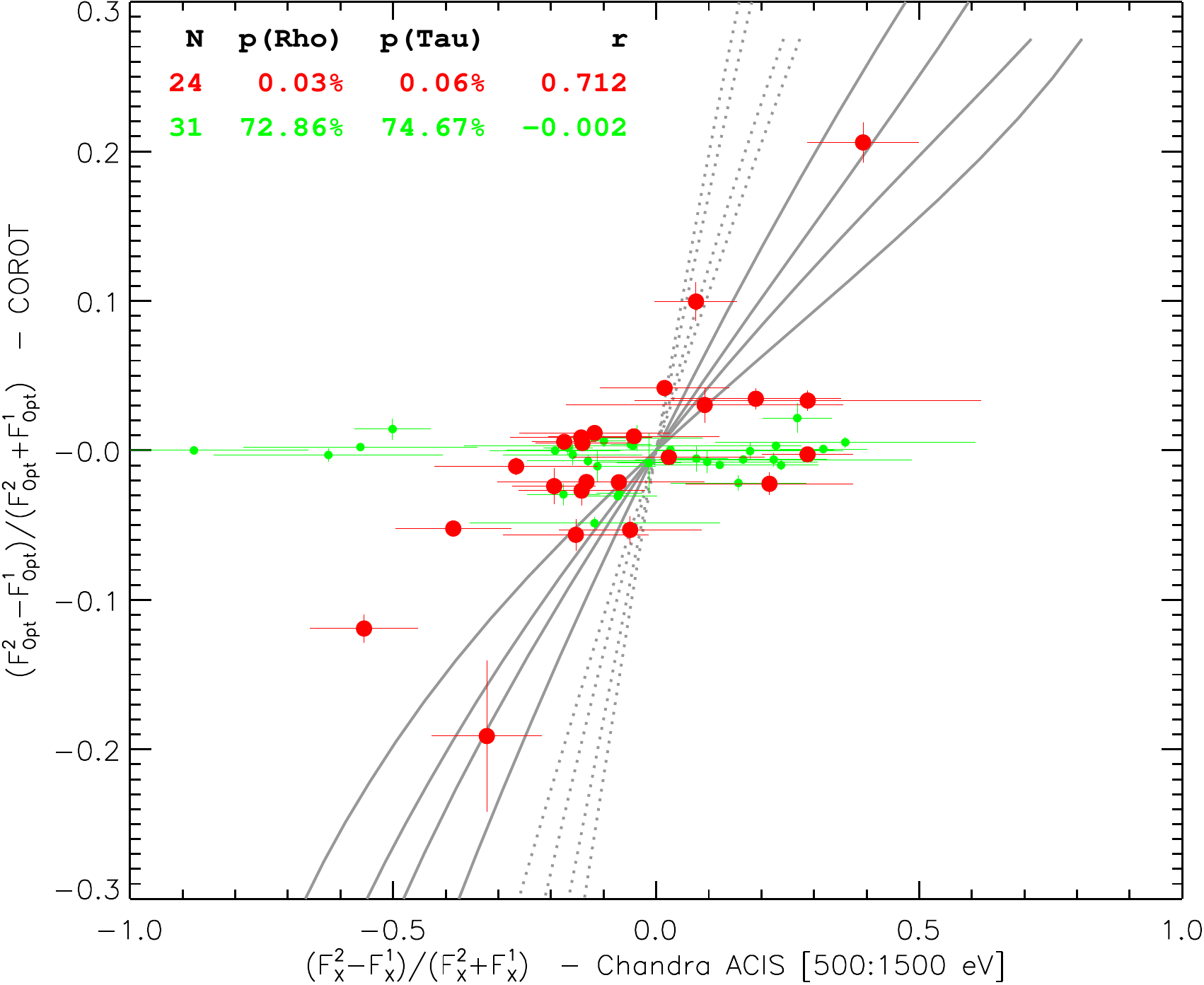}
\includegraphics[width=9.15cm]{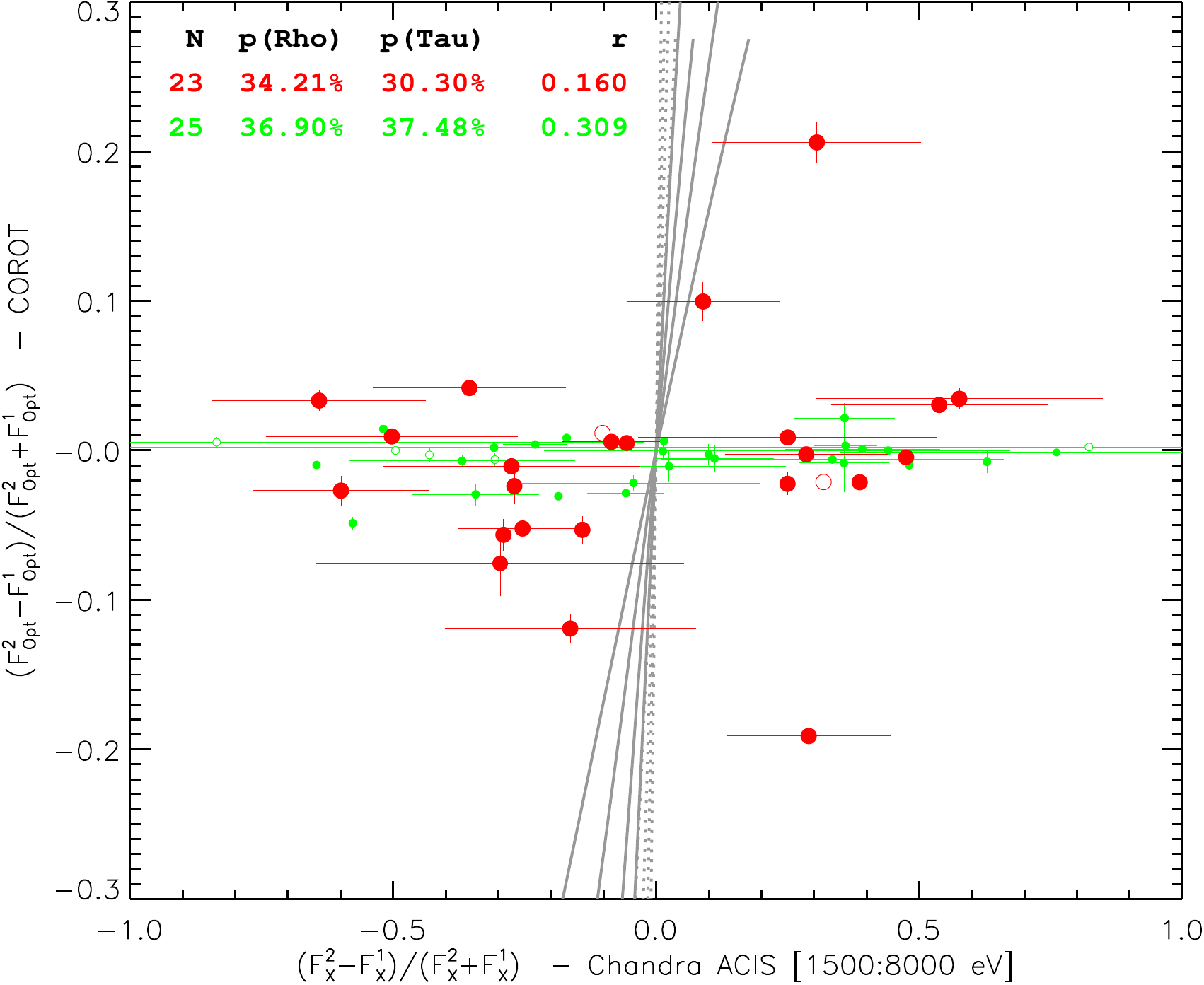}
  \caption{Optical vs. X-ray flux variations between the two $\sim$30\,ksec {\em Chandra} observations separated by $\sim$16 days. The left- and righthand panels refer to the {\em soft} and {\em hard} X-ray bands, respectively (500:1500\,eV and 1500:8000\,eV). We plot on each axis the differences of the two optical or X-ray fluxes, averaged over the {\em Chandra} exposures, divided by the sum of the same quantities (i.e. half of the fractional variations with respect to the mean fluxes). CTTS and WTTS are indicated with large (red) and small (green) symbols, respectively. Filled circles refer to stars for which we collected at least 5 X-ray photons, in the relevant energy band, in either of the {\em Chandra} exposures (\S \ref{sect:data}). Empty circles in the righthand panel refer to stars that satisfy this condition in the soft band, but not in the hard band.
  Error bars on the X-ray variations are propagated from the Poisson uncertainties on the observed ACIS countrates. Those on the optical variations reflect instead the variance of the binned CoRoT lightcurves within the $\sim30$\,ksec observing windows. In each panel, the number of CTTS and WTTS satisfying our selection criterion on the X-ray statistics are reported in the upper-left corner, followed by results of statistical tests conducted on these samples: the null probabilities given by the Spearman's ($\rho$) and Kendalls's ($\tau$) rank correlation tests and the correlation coefficient, $r$. A significant correlation is observed only for the soft X-ray flux of CTTSs.
  The gray lines indicate the effect of absorption on an intrinsically constant source with average A$_{\rm V}$=1.0, assuming, respectively, a standard ``interstellar'' gas/dust ratio (dotted lines), and a 5 times higher gas/dust ratio (solid lines). The four lines in each set refer to the combinations of two plasma temperatures, 0.6 and 5.0\,keV, and two different assumptions for the spectrum-dependent extinction law in the wide CoRoT band: $\rm A_{CoRoT}/A_{V}=1.0$ and $\rm A_{CoRoT}/A_{V}=\rm A_{I}/A_{V}$=0.61.}
\label{fig}
\end{figure*}

Figure\,\ref{fig} shows scatter plots between the optical and X-ray
fractional variability of CTTSs and WTTSs, separately for the soft and
hard X-ray bands. Sample sizes, the results of the Spearman's ($\rho$)
and Kendalls's ($\tau$) rank correlation tests, and the correlation
coefficients $r$ are reported in the two panels. The optical and soft
X-ray variability of CTTSs is significantly correlated (null
probabilities 0.03-0.06\%). No such statistical evidence is visible
for WTTs, where the X-ray and optical variations appear to be
uncorrelated. Also, the amplitude of the optical variability is
significantly higher for CTTSs than for WTTSs. An inspection of the
lightcurves shows that regular rotational modulation dominates in
$\sim$80\% of the WTTSs, while $\sim$90\% of the CTTSs show irregular or
AA Tau-like variability \citep[cf.][]{ale10}.

No correlation is evident in the hard band. While the lower statistics
make it harder to detect a correlation, the scatter of points in all
quadrants point to a different physical mechanism driving variability in
the soft and in the hard X-ray bands, with only the soft X-ray
variability linked with the optical one. Given the error bars
and the limited sample sizes, it is unfortunately impossible to
determine whether the correlation for the soft X-ray band is common to
{\em all} the CTTS in our sample, or if distinct behaviors characterize
the stars with larger X-ray/optical variability and those with smaller
amplitudes. However, the correlation is probably not driven just by an
handful of stars: excluding the four most optically variable stars, the
correlation tests still indicate a likely correlation, with null
probabilities of 2.8-3.2\%. The extension of the correlation to the low-variability stars is even more striking considering that whatever
drives variability in WTTSs (e.g., dark spots in the optical and active
region evolution/rotational modulation in the X-rays) is likely at work
also on CTTSs, thus diluting any CTTS-specific effect at low amplitudes.

\section{Discussion}

The observation of correlated optical and soft X-ray variability for CTTSs in NGC~2264 is at odds with the negative results reported by \citet{gro07}, for AA Tau,  and \citet{sta06} for a sample of Orion stars. The different quality of the optical lightcurves and different timescales probed by these studies might explain the apparent contradiction.

Our result points either to $i$) a single physical mechanism modulating the optical and soft X-ray emission or $ii$) the emitting regions being obscured by the same material and thus the emission being absorbed in a correlated fashion. 
In principle, the observed X-ray spectra would vary differently in these two scenarios. The limited statistics of the low-resolution X-ray spectra obtained in our {\em Chandra} observation and the known degeneracy between temperature and absorption when fitting these spectra with models made the spectral analysis inconclusive.

If both the varying optical emissions and the varying soft X-ray emission were to come from the accretion shock, one could invoke a mechanism of type $i$) above: modulation in the mass accretion rate. To explain the observed variability range (approximately up to $\pm 40\%$), this would require the CoRoT {\em broad-band} optical emission being dominated by accretion luminosity, a condition that is not verified for most CTTSs \citep[cf.][]{gul98}. At the same time, the soft X-ray spectrum should be dominated by emission from the shock region, with little or no coronal contribution, to explain the approximately $\pm 100\%$ variability range. However, all these stars show evidence of significant coronal emission (as shown by the thermal emission in the harder band), and only in the peculiar CTTS TW\,Hya is the X-ray emission dominated by the shock region. In order to explain the correlated variability we might also speculate that the {\em coronal} emission is directly correlated to accretion. This hypothesis is, however, not supported by observations that, if anything, indicate an inverse relation between accretion and coronal activity \citep[e.g.][]{pre05a}. It thus appears unlikely that the large variations in the soft X-ray and in the optical fluxes can be attributed to variability in the accretion shock region.

The second class of mechanisms assume shading of the optical emission region (photosphere) and of the X-ray emission region (corona) by the same (or correlated) circumstellar material. Rotationally-modulated shading of the optical emission (photospheric and from the accretion spot) is observed in AA\,Tau, and similar variations in the optical lightcurve are observed in several CTTSs in NGC\,2264. Under the assumption that the circumstellar material covers a similar fraction of coronal and photospheric material, the correlated flux variability may be explained quite naturally by time-variable shading. 

The differences in the mechanisms leading to the absorption of X-ray and optical photons give additional diagnostic power to the observed correlation. X-ray photons are absorbed mainly by material in the gas phase while optical photons are absorbed mainly by dust grains. In the hypothesis that the optical emitting region (the photosphere and the accretion spot) and the X-ray emitting one (the corona) are, from the point of view of the shading material, similar in location and size (i.e. that they are at all times subject to comparable shading), the shape of the diagram in Fig.~\ref{fig} will be determined by the extinction laws in the optical and X-rays and by the gas-to-dust ratio in the shading material. We computed theoretical loci in Fig.~\ref{fig}, starting from the ``standard'' relation between optical extinction, A$_{\rm V}$, and N$_{\rm H}$, the equivalent hydrogen column density with which X-ray absorption is usually parametrized: N$_{\rm H}$=$1.6\times 10^{21}\rm A_V$\,cm$^{-2}$ \citep{vuo03}, valid for an interstellar gas-to-dust ratio. The effect of non-standard gas-to-dust ratio is easily mimicked by multiplying the above  N$_{\rm H}$ -  A$_{\rm V}$ relation by a gas-enrichment factor.  
The theoretical loci in Fig.\,\ref{fig} depend weakly on the incident X-ray spectrum (that we modeled as a single-temperature thermal plasma) and on the photospheric temperature, and we have assumed a plausible range for both parameters, computing ``families of loci'' as a function of the gas-enrichment factor. As shown in Fig.\,\ref{fig}, a ``standard/interstellar'' gas-to-dust ratio would require the X-rays variations to be, in relation to the optical variations, significantly smaller than observed.

One natural way of explaining the observed correlation slope is to assume an absorbing medium depleted in dust grains (thus primarily absorbing X-rays). The circumstellar material in the immediate vicinity of the star, inside the circumstellar inner rim, including the accretion streams, is indeed expected to be heavily gas-depleted \citep[e.g.][]{kam09,ise09} and an enhanced absorption of X-rays from CTTSs has actually been reported \citep[e.g.][]{gun08}. While we have not attempted to formally fit the data to derive the average degree of dust depletion in the absorbing medium, as shown in Fig.~\ref{fig}, the locii corresponding to a five-fold depletion factor in the dust content with respect to the ``standard'' interstellar value do provide a good explanation for the observed correlation.

An alternative explanation could be built by assuming a standard gas-to-dust ratio together with a more substantial shading of the X-ray emitting region with respect to the photosphere and the accretion spot. It is easy to imagine such situations: an equatorial corona with a disk warp only eclipsing the low-latitude regions of the photosphere would be a possible example, as would emission from a low-latitude flux tube extending from the photosphere to the inner disk, when the flux tube is pointing toward the observer. However, to explain a general correlation would require that these specific geometrical conditions be realized for all CTTSs in our sample, or at least for those with large variability amplitudes that determine the statistical correlation. This would require ad hoc assumptions about the shape of the warped disk (assuming an AA\,Tau-like configuration), the location of the corona, and the relative inclination of the systems as observed. Again, while this is not impossible to envisage, a higher gas-to-dust ratio in the absorbing medium provides a much more natural and simpler explanation.

The observed correlation and the amplitude of the variability imply that a significant fraction of the X-ray emission from CTTSs is affected by shading and obscuration. If this obscuration is sufficiently optically thick, it may be impossible to recognize it and to determine its amount with the usual X-ray spectral analysis techniques. Since part of the corona would remain unaccounted for, the shading thus also provides a natural explanation for the lower X-ray luminosity of CTTSs with respect to WTTSs for stars of the same mass, as well as for the wider range in observed X-ray luminosities \citep[cf.][]{gre07}.

If our favored interpretation of the observed correlation is correct, the observed X-ray variability, on the 16-day timescale probed by our observations, is for a large fraction not intrinsic to the emitting source but rather caused by variations in the absorbing material. Because of this, future simultaneous observations of CTTSs in the optical and soft X-rays, spanning a range of time scales comparable to the rotational period of the stars, can provide unique diagnostics of their circumstellar environment.

\begin{acknowledgements}
We thank Costanza Argiroffi and Salvo Sciortino for discussions that helped shape our conclusions, and the anonymous referee for useful suggestions on how to improve this work.
\end{acknowledgements}

\bibliographystyle{aa} %aa.bst\
\bibliography{bibtex.bib}

\begin{thebibliography}{30}
\expandafter\ifx\csname natexlab\endcsname\relax\def\natexlab#1{#1}\fi

\bibitem[{{Alencar} {et~al.}(2010){Alencar}, {Teixeira}, {Guimaraes},
  {McGinnis}, {Gameiro}, {Bouvier}, {Aigrain}, {Flaccomio}, \&
  {Favata}}]{ale10}
{Alencar}, S.~H.~P., {Teixeira}, P.~S., {Guimaraes}, M.~M., {et~al.} 2010,
  \aap, in press

\bibitem[{{Bouvier} {et~al.}(2007){Bouvier}, {Alencar}, {Boutelier},
  {Dougados}, {Balog}, {Grankin}, {Hodgkin}, {Ibrahimov}, {Kun}, {Magakian}, \&
  {Pinte}}]{bou07}
{Bouvier}, J., {Alencar}, S.~H.~P., {Boutelier}, T., {et~al.} 2007, \aap, 463,
  1017

\bibitem[{{Bouvier} {et~al.}(1999){Bouvier}, {Chelli}, {Allain}, {Carrasco},
  {Costero}, {Cruz-Gonzalez}, {Dougados}, {Fern{\'a}ndez}, {Mart{\'{\i}}n},
  {M{\'e}nard}, {Mennessier}, {Mujica}, {Recillas}, {Salas}, {Schmidt}, \&
  {Wichmann}}]{bou99}
{Bouvier}, J., {Chelli}, A., {Allain}, S., {et~al.} 1999, \aap, 349, 619

\bibitem[{{Bouvier} {et~al.}(2003){Bouvier}, {Grankin}, {Alencar}, {Dougados},
  {Fern{\'a}ndez}, {Basri}, {Batalha}, {Guenther}, {Ibrahimov}, {Magakian},
  {Melnikov}, {Petrov}, {Rud}, \& {Zapatero Osorio}}]{bou03}
{Bouvier}, J., {Grankin}, K.~N., {Alencar}, S.~H.~P., {et~al.} 2003, \aap, 409,
  169

\bibitem[{{Broos} {et~al.}(2010){Broos}, {Townsley}, {Feigelson}, {Getman},
  {Bauer}, \& {Garmire}}]{bro10}
{Broos}, P.~S., {Townsley}, L.~K., {Feigelson}, E.~D., {et~al.} 2010, ArXiv
  e-prints

\bibitem[{{Dahm} \& {Simon}(2005)}]{dah05}
{Dahm}, S.~E. \& {Simon}, T. 2005, \aj, 129, 829

\bibitem[{{Damiani} {et~al.}(1997){Damiani}, {Maggio}, {Micela}, \&
  {Sciortino}}]{dam97}
{Damiani}, F., {Maggio}, A., {Micela}, G., \& {Sciortino}, S. 1997, \apj, 483,
  350

\bibitem[{{F{\H u}r{\'e}sz} {et~al.}(2006){F{\H u}r{\'e}sz}, {Hartmann},
  {Szentgyorgyi}, {Ridge}, {Rebull}, {Stauffer}, {Latham}, {Conroy},
  {Fabricant}, \& {Roll}}]{fur06}
{F{\H u}r{\'e}sz}, G., {Hartmann}, L.~W., {Szentgyorgyi}, A.~H., {et~al.} 2006,
  \apj, 648, 1090

\bibitem[{{Flaccomio} {et~al.}(2003){Flaccomio}, {Damiani}, {Micela},
  {Sciortino}, {Harnden}, {Murray}, \& {Wolk}}]{fla03b}
{Flaccomio}, E., {Damiani}, F., {Micela}, G., {et~al.} 2003, \apj, 582, 398

\bibitem[{{Flaccomio} {et~al.}(2006){Flaccomio}, {Micela}, \&
  {Sciortino}}]{fla06}
{Flaccomio}, E., {Micela}, G., \& {Sciortino}, S. 2006, \aap, 455, 903

\bibitem[{{Grankin} {et~al.}(2008){Grankin}, {Bouvier}, {Herbst}, \&
  {Melnikov}}]{gra08}
{Grankin}, K.~N., {Bouvier}, J., {Herbst}, W., \& {Melnikov}, S.~Y. 2008, \aap,
  479, 827

\bibitem[{{Grankin} {et~al.}(2007){Grankin}, {Melnikov}, {Bouvier}, {Herbst},
  \& {Shevchenko}}]{gra07}
{Grankin}, K.~N., {Melnikov}, S.~Y., {Bouvier}, J., {Herbst}, W., \&
  {Shevchenko}, V.~S. 2007, \aap, 461, 183

\bibitem[{{Gregory} {et~al.}(2007){Gregory}, {Wood}, \& {Jardine}}]{gre07}
{Gregory}, S.~G., {Wood}, K., \& {Jardine}, M. 2007, \mnras, 379, L35

\bibitem[{{Grosso} {et~al.}(2007){Grosso}, {Bouvier}, {Montmerle},
  {Fern{\'a}ndez}, {Grankin}, \& {Zapatero Osorio}}]{gro07}
{Grosso}, N., {Bouvier}, J., {Montmerle}, T., {et~al.} 2007, \aap, 475, 607

\bibitem[{{G{\"u}del} \& {Telleschi}(2007)}]{gue07b}
{G{\"u}del}, M. \& {Telleschi}, A. 2007, \aap, 474, L25

\bibitem[{{Gullbring} {et~al.}(1998){Gullbring}, {Hartmann}, {Briceno}, \&
  {Calvet}}]{gul98}
{Gullbring}, E., {Hartmann}, L., {Briceno}, C., \& {Calvet}, N. 1998, \apj,
  492, 323

\bibitem[{{G{\"u}nther} \& {Schmitt}(2008)}]{gun08}
{G{\"u}nther}, H.~M. \& {Schmitt}, J.~H.~M.~M. 2008, \aap, 481, 735

\bibitem[{{Imanishi} {et~al.}(2001){Imanishi}, {Koyama}, \& {Tsuboi}}]{ima01}
{Imanishi}, K., {Koyama}, K., \& {Tsuboi}, Y. 2001, \apj, 557, 747

\bibitem[{{Isella} {et~al.}(2009){Isella}, {Carpenter}, \& {Sargent}}]{ise09}
{Isella}, A., {Carpenter}, J.~M., \& {Sargent}, A.~I. 2009, \apj, 701, 260

\bibitem[{{Joy}(1945)}]{joy45}
{Joy}, A.~H. 1945, \apj, 102, 168

\bibitem[{{Kama} {et~al.}(2009){Kama}, {Min}, \& {Dominik}}]{kam09}
{Kama}, M., {Min}, M., \& {Dominik}, C. 2009, \aap, 506, 1199

\bibitem[{{Lamm} {et~al.}(2004){Lamm}, {Bailer-Jones}, {Mundt}, {Herbst}, \&
  {Scholz}}]{lam04}
{Lamm}, M.~H., {Bailer-Jones}, C.~A.~L., {Mundt}, R., {Herbst}, W., \&
  {Scholz}, A. 2004, \aap, 417, 557

\bibitem[{{M{\'e}nard} {et~al.}(2003){M{\'e}nard}, {Bouvier}, {Dougados},
  {Mel'nikov}, \& {Grankin}}]{men03}
{M{\'e}nard}, F., {Bouvier}, J., {Dougados}, C., {Mel'nikov}, S.~Y., \&
  {Grankin}, K.~N. 2003, \aap, 409, 163

\bibitem[{{Preibisch} {et~al.}(2005){Preibisch}, {Kim}, {Favata}, {Feigelson},
  {Flaccomio}, {Getman}, {Micela}, {Sciortino}, {Stassun}, {Stelzer}, \&
  {Zinnecker}}]{pre05a}
{Preibisch}, T., {Kim}, Y.-C., {Favata}, F., {et~al.} 2005, \apjs, 160, 401

\bibitem[{{Stassun} {et~al.}(2007){Stassun}, {van den Berg}, \&
  {Feigelson}}]{sta07}
{Stassun}, K.~G., {van den Berg}, M., \& {Feigelson}, E. 2007, \apj, 660, 704

\bibitem[{{Stassun} {et~al.}(2006){Stassun}, {van den Berg}, {Feigelson}, \&
  {Flaccomio}}]{sta06}
{Stassun}, K.~G., {van den Berg}, M., {Feigelson}, E., \& {Flaccomio}, E. 2006,
  \apj, 649, 914

\bibitem[{{Sung} {et~al.}(2008){Sung}, {Bessell}, {Chun}, {Karimov}, \&
  {Ibrahimov}}]{sun08}
{Sung}, H., {Bessell}, M.~S., {Chun}, M., {Karimov}, R., \& {Ibrahimov}, M.
  2008, \aj, 135, 441

\bibitem[{{Sung} {et~al.}(2009){Sung}, {Stauffer}, \& {Bessell}}]{sun09}
{Sung}, H., {Stauffer}, J.~R., \& {Bessell}, M.~S. 2009, \aj, 138, 1116

\bibitem[{{Vuong} {et~al.}(2003){Vuong}, {Montmerle}, {Grosso}, {Feigelson},
  {Verstraete}, \& {Ozawa}}]{vuo03}
{Vuong}, M.~H., {Montmerle}, T., {Grosso}, N., {et~al.} 2003, \aap, 408, 581

\bibitem[{{Walker}(1956)}]{wal56}
{Walker}, M.~F. 1956, \apjs, 2, 365

\end{thebibliography}

\end{document}